\documentclass{article}

\usepackage[numbers,comma]{natbib}

\usepackage[preprint]{neurips_2025}

\usepackage[utf8]{inputenc} 
\usepackage[T1]{fontenc}    
\usepackage{hyperref}       
\usepackage{url}            
\usepackage{booktabs}       
\usepackage{amsfonts}       
\usepackage{nicefrac}       
\usepackage{microtype}      
\usepackage{xcolor}         

\usepackage{amsmath}
\usepackage{multirow}
\usepackage[normalem]{ulem}
\useunder{\uline}{\ul}{}
\usepackage{amssymb}
\usepackage{dsfont}
\usepackage{caption}
\usepackage{diagbox}
\usepackage{graphicx}
\usepackage{wrapfig}
\usepackage{dutchcal}
\usepackage{color}

\usepackage{bbding}
\usepackage{pifont}
\usepackage{wasysym}
\usepackage{amssymb}
\usepackage{color, colortbl}
\usepackage{csquotes}
\usepackage{enumitem}
\usepackage{longtable}

\usepackage{comment}

\title{VRAgent-R1: Boosting Video Recommendation with MLLM-based Agents via Reinforcement Learning}

\author{
Chen Siran\textsuperscript{1,2}, 
Chen Boyu\textsuperscript{1,2},
Yu Chenyun\textsuperscript{3},
Luo Yuxiao\textsuperscript{1},
Yi Ouyang\textsuperscript{2},
\\
\textbf{Cheng Lei\textsuperscript{2},}
\textbf{Zhuo Chengxiang\textsuperscript{2},}
\textbf{Li Zang\textsuperscript{2},}
\textbf{Wang Yali\textsuperscript{1,4}} \\
\\
  \textsuperscript{1}SIAT@MMLab, 
  \textsuperscript{2} Platform and Content Group, Tencent,\\
  \textsuperscript{3}Sun Yat-sen University,
  \textsuperscript{4}Shanghai AILab,
}

\begin{document}

\maketitle

\begin{abstract}

Owing to powerful natural language processing and generative capabilities, large language model (LLM) agents have emerged as a promising solution for enhancing recommendation systems via user simulation. 
However, in the realm of video recommendation, existing studies predominantly resort to prompt-based simulation using frozen LLMs and encounter the intricate challenge of multimodal content understanding. This frequently results in suboptimal item modeling and user preference learning, thereby ultimately constraining recommendation performance.
To address these challenges, we introduce VRAgent-R1, a novel agent-based paradigm that incorporates human-like intelligence in user simulation. Specifically, VRAgent-R1 comprises two distinct agents: the Item Perception (IP) Agent and the User Simulation (US) Agent, designed for interactive user-item modeling.
Firstly, the IP Agent emulates human-like progressive thinking based on MLLMs, effectively capturing hidden recommendation semantics in videos. With a more comprehensive multimodal content understanding provided by the IP Agent, the video recommendation system is equipped to provide higher-quality candidate items.
%
Subsequently, the US Agent refines the recommended video sets based on in-depth chain-of-thought (CoT) reasoning and achieves better alignment with real user preferences through reinforcement learning.
Experimental results on a large-scale video recommendation benchmark have demonstrated the effectiveness of our proposed VRAgent-R1 method,
e.g., the IP Agent achieves a 6.0\% improvement in NDCG@10 on the MicroLens-100k dataset, while the US Agent shows approximately 45.0\% higher accuracy in user decision simulation compared to state-of-the-art baselines.

\end{abstract}

\section{Introduction}

With the widespread adoption of online multimedia services, particularly the booming popularity of short video platforms like TikTok and Kuaishou, Multimodal Recommendation Systems (MRS) have attracted considerable attention from both academia and industry. 
Unlike conventional recommendation technologies~\cite{huang2015tencentrec,zheng2018spectral,ying2018graph,yang2020mixed,yuan2020parameter} that rely on ID-based user/item representation learning, 
MRS focuses on effectively integrating data associated with items from diverse modalities, such as text, images, videos, and audio,
which aims to offer more personalized recommendations by deep semantic understanding of item and user characteristics.
%
%
%
However, in video recommendations, 
existing approaches predominantly construct item representations based on text and images due to challenges in computational efficiency and processing heterogeneous information across modalities,
the full potential of multimodal information in video recommendation systems remains largely unexplored.

Benefiting from the significant advancements in Multimodal Large Language Models (MLLMs),
recent studies tend to utilize their strengths and potential to boost multimodal recommendation systems.
On one hand, some approaches~\cite{zhang2024notellm,chen2024hllm,luo2024qarm,sun2019bert4rec,ren2024representation,lee2024star,jia2025learn,song2024precise} leverage the pre-trained MLLMs to directly convert each item's multimodal information into a single embedding.
However, fine-tuning MLLMs and achieving semantic alignment for recommendation requires abundant high-quality interaction data and computational resources.
On the other hand, some methods~\cite{zhang2025llm,MLLM-MSR,bao2023tallrec,zhang2024usimagent} directly apply LLMs or MLLMs to recommendation tasks, transforming the recommendation into a language generation problem. 
For instance, by analyzing user interaction history, item descriptions, or conversational context, LLMs perform in-depth inference to generate recommendation results.
While effective in addressing the data sparsity issue, 
these methods often face challenges such as limited input length, computational inefficiency, and hallucinations, 
making them unsuitable for large-scale recommendations.
Recently, growing attention has been paid to using LLM-based agents to enhance recommendation systems' personalization and intelligence (e.g., user profiling, simulating interactions, improving satisfaction). 
However, existing methods~\cite{zhang2024agent4rec,zhang2025llm,xiang2024simuser} mostly use frozen LLMs, with the knowledge gap between vertical domains and LLMs/MLLMs limiting their adaptability and effectiveness.





\begin{figure}[t]
    \centering
    \setlength{\abovecaptionskip}{0.1cm}
    \includegraphics[width=1.0\linewidth]{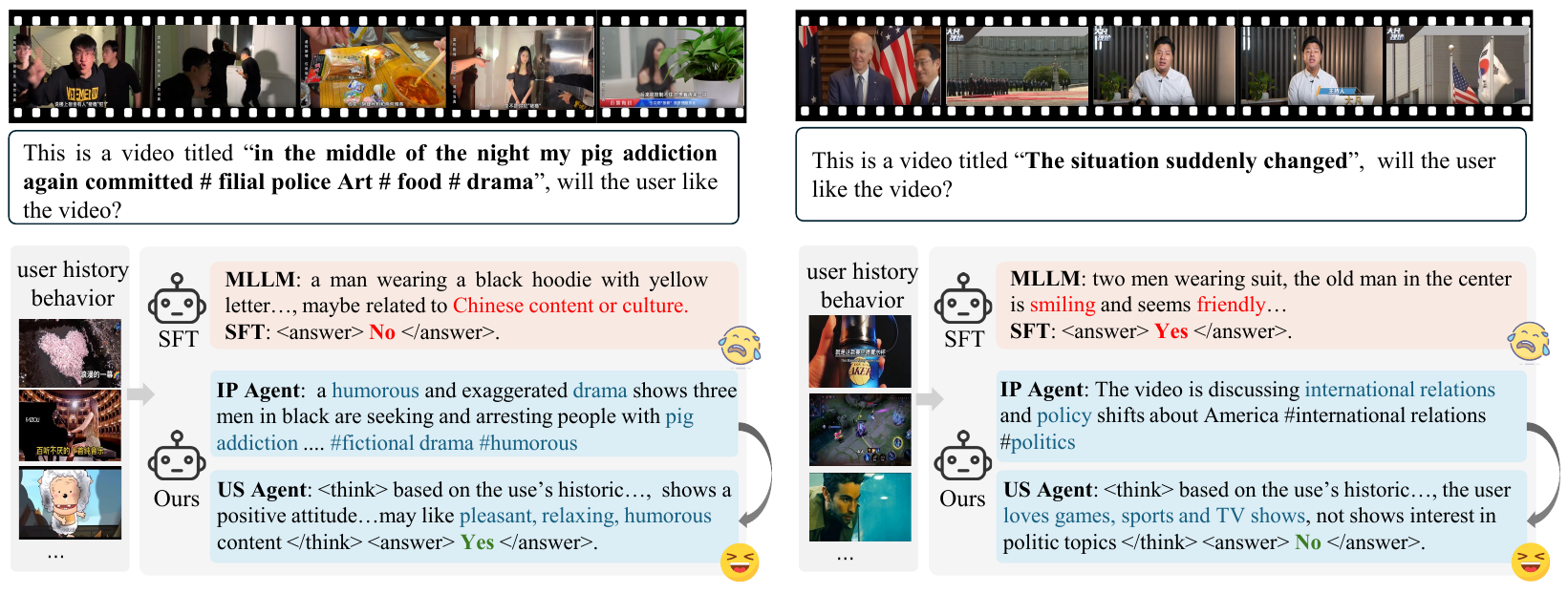}
    \caption{\textbf{Qualitative examples of VRAgent-R1 for Video Recommendation.} Our VRAgent-R1 stands out from previous supervised fine-tuning of MLLM, which fails to give the correct prediction due to the lack of understanding of video items and deep thinking on user status.}
    \vspace{-0.3cm}
    \label{fig:vis}
    \vspace{-0.5cm}
\end{figure}

Based on the above discussion, we propose to use LLMs/MLLMs to attain enhanced multimodal content comprehension
and simulate user decision-making.
Our objective is to optimize representation learning and human-centric recommendation outcomes, so as to align with real user preferences.
In order to realize satisfactory MLLM-based user simulation for video recommendation, we are required to address the following challenges.
\textbf{1) How to discriminately exploit the recommendation-relevant semantics hidden in video items?}
Most video recommendation systems heavily rely on understanding video items, which presents two major difficulties.
Firstly, previous methods ignore the temporal relation within video frames and fail to identify key information among numerous contents.
For example,
some methods~\cite{he2016vista,ni2023content} directly use randomly selected, pre-processed visual features,
while MLLM-MSR~\cite{MLLM-MSR} only takes the first image for video modeling, all of which fail to effectively utilize the characteristics of videos.
Secondly,
previous methods process each modality independently and conduct late feature fusion, which faces the problem of modality competition~\cite{shang2023enhancing,liu2024survey}, even leading to inferior performance compared to using a single modality~\cite{zhou2023comprehensive}.
Additionally, the lack of in-depth semantic interaction between raw modalities may also cause a misunderstanding of the high-level semantics of the video. As shown in Fig. \ref{fig:MLLM} (a), (b),
these methods merely obtain superficial modality embeddings while failing to identify the political topics expressed in the video.
\textbf{2) How to effectively simulate user behavior that mirrors human deep thinking?}
LLMs show great promise for user simulation in the recommendation systems~\cite{zhang2024agent4rec,zhang2024usimagent,zhang2025llm,wang2025user} due to their excellent linguistic understanding capabilities.
However,
existing LLMs struggle with processing sequential multimodal inputs, which prevents direct modeling of multimodal user behavior sequences. 
Moreover, most user simulators~\cite{zhang2024agent4rec,zhang2025llm,zhang2024usimagent,xiang2024simuser} primarily rely on prompt engineering to instruct frozen LLMs to generate responses without feedback,
potentially leading to discrepancies between the agent and real user behavior when the model cannot be optimized.
Some methods~\cite{bao2023tallrec,MLLM-MSR} attempt to fine-tune the large model, 
but simple fine-tuning only yields binary `Yes' or `No' judgments without deep analysis of the user's status, 
thus limiting the model's generalizability as shown in Fig. \ref{fig:vis}.

To address the aforementioned challenges,
we introduce VRAgent-R1, a novel agent-based paradigm for video recommendation.
Unlike prior approaches built on frozen LLMs, VRAgent-R1 effectively mimics the human-like thinking for recommendation by leveraging multimodal collaborative understanding and reinforcement fine-tuning (RFT) on user simulation.
Specifically, it consists of two distinct agents:
the Item Perception (IP) Agent and the User Simulation (US) Agent.
As illustrated in Fig. \ref{fig:intro},
The IP Agent aims to establish comprehensive multimodal content understanding for videos to improve item modeling 
through multi-round, in-depth semantic interaction with the MLLM.
This process enables it to progressively discover key video content and effectively extract recommendation-relevant semantics from the items.
Subsequently, the semantic summarization of videos generated by the IP Agent is used to optimize the fundamental video recommendation model via feature augmentation. This not only enhances the recommendation model but also facilitates the US Agent to capture user preferences and predict the next item based on historical interactions.
The US Agent focuses on deep user behavior simulation and provides proxy feedback to refine the candidate set provided by the video recommendation system.
To align user simulation with real decision-making,
reinforcement learning is leveraged to enable the model to analyze historical user behavior (watched videos and comments) and comprehensively summarize user status with Chain-of-Thought (CoT) reasoning.
Additionally, we design a reward mechanism associated with the user's actual final behavior and update the fundamental LLM through policy optimization, i.e., GRPO~\cite{shao2024deepseekmath}.
Through the collaboration of IP and US agents, VRAgent-R1 effectively boosts video recommendation performance with step-by-step human-like thinking.
The main contributions of this work can be summarized as follows:

\begin{figure}[t] 
    \centering
    \setlength{\abovecaptionskip}{0.1cm}
    \includegraphics[width=\linewidth]{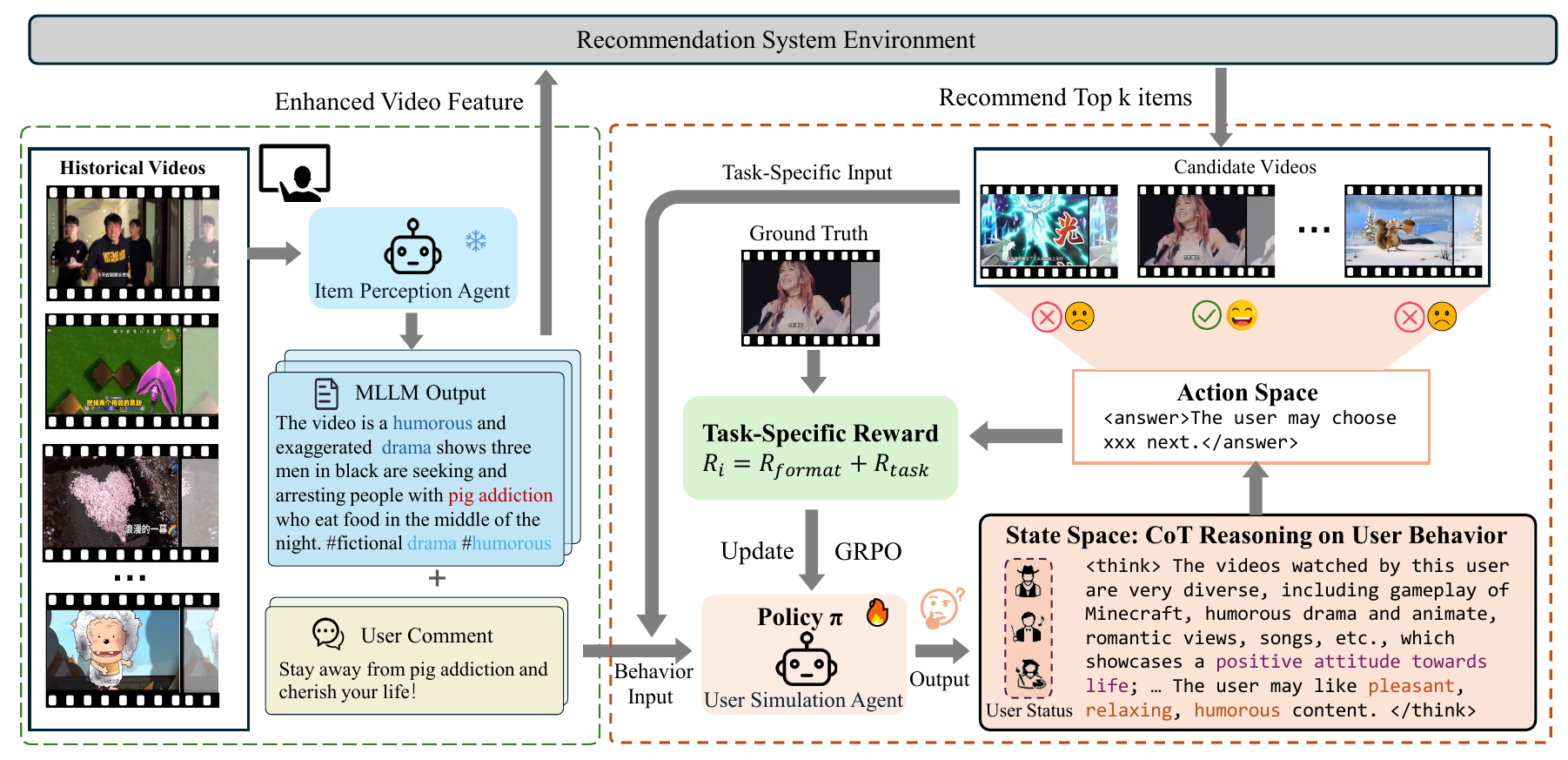}
    \caption{\textbf{Overview of our VRAgent-R1 framework.} 
    We propose a framework with two novel agents for better video recommendation. The IP Agent conducts collaborative multimodal understanding to obtain enhanced video features for the recommendation system and the US Agent. Meanwhile, the US Agent simulates user behavior via deep CoT reasoning based on user status. By reinforcement learning with actual behavior rewards, VRAgent-R1 achieves superior simulation performance and helps improve the recommendation accuracy.
    }
    \vspace{-0.3cm}
    \label{fig:intro}
    \vspace{-0.3cm}
\end{figure}



%
\vspace{-0.3cm}

\begin{itemize}[itemsep= 1pt,topsep = 5 pt,leftmargin = 10pt]
\item
We propose VRAgent-R1, a novel agent-based framework designed to assist video recommendations from a user-centric perspective, which exhibits human-like intelligence for interpretable recommendations. By incorporating a user-like understanding, this framework significantly enhances the performance of recommendation system, demonstrating the effectiveness of the pipeline.
\item 
Our IP Agent achieves more comprehensive multimodal understanding of videos by flexibly conducting in-depth semantic interactions between textual and visual contents. Furthermore, our US Agent is the first to use RFT for LLM-based user simulation, achieving more accurate simulation performance through deep thinking on user status with little training data.
\item 
Extensive experiments demonstrate that the IP Agent significantly enhances the performance of existing video recommendation approaches, achieving a 4.3\% improvement in HR@10 and a 6.0\% improvement in NDCG@10 on the MicroLens-100k dataset~\cite{ni2023content}.
Meanwhile, the US Agent outperforms commercial models such as GPT-4o~\cite{gpt4o} and SFT methods,
attaining an accuracy of 64.1\% in user simulation for next video selection, which represents a 45.0\% enhancement over the SFT baseline.
Moreover, simulated user feedback can further boost the recommendation accuracy by reranking the candidate item set generated from the recommendation system.
\end{itemize}

\section{Related Work}

\vspace{-0.1cm}
\textbf{LLMs/MLLMs for Multi-Modal Recommendation.}
LLMs and MLLMs have performed a profound impact on their integration into current recommendation systems. 
Existing methods utilizing LLMs can be broadly categorized into implicit and explicit applications.
The implicit methods~\cite{sun2019bert4rec,ren2024representation,lee2024star,jia2025learn,song2024precise,zhang2024notellm,chen2025super,chen2024percept} directly utilize the pre-trained structure or parameters of large models to convert user and item information into embeddings.
For example,
LEARN~\cite{jia2025learn} integrates key attributes such as the title, description, and brand of each item into a predefined prompt,
and then uses the last layer feature of the LLM as the item embedding.
NoteLLM-2~\cite{zhang2024notellm} employs an MLLM with end-to-end fine-tuning to fuse multimodal information as the item embedding. 
Explicit methods~\cite{zhang2025llm,zhang2024usimagent,hou2024large,MLLM-MSR,bao2023tallrec} involve using the reasoning ability of MLLMs to expand item information and analyze user profiles or intentions, ultimately generating textual summarization to aid recommendations.
For example, MLLM-MSR~\cite{MLLM-MSR} uses MLLMs to generate detailed image captions and summarize user preferences for further recommendations.
However, research on video recommendation with minute-level visual content is relatively scarce compared to text-based and image-based recommendations, and we are pioneers in using MLLMs for video recommendation.

\textbf{LLM-based User Simulator.}
Considering the powerful semantic understanding and reasoning abilities, many works utilize LLM to assist user inference simulations~\cite{wang2023rethinking,zhang2024usimagent,corecco2024suber,zhang2024agent4rec,zhang2025llm}.
For instance, iEvaLM~\cite{wang2023rethinking} explores two types of interaction within a conversational recommendation benchmark: attribute-based question answering and free-form chit-chat using ChatGPT~\cite{achiam2023gpt}.
To simulate user search behavior, USimAgent~\cite{zhang2024usimagent}  prompts an LLM agent to construct complete search sessions, including querying, clicking, and stopping behaviors, according to specific search tasks.
Agent4Rec~\cite{zhang2024agent4rec} initializes LLMs as agents with unique user profiles that encompass tastes and social traits to simulate more realistic user behaviors, thereby reflecting user preferences and social characteristics.
Additionally, LLM\_Simulator~\cite{zhang2025llm} simulates user preferences by matching the positive and negative attributes of items with user preferences generated by the LLM to determine whether a user would like an item. 
However, previous LLM-based user simulation approaches have relied on frozen LLMs, and using them solely through prompting would risk discrepancies with real user behavior and potential hallucinations~\cite{zhang2024agent4rec}.

\textbf{Training LLMs/MLLMs with RL.}
With the success of DeepSeek-R1~\cite{guo2025deepseek}, reinforcement learning (RL) has demonstrated its remarkable ability to enhance the logical reasoning capabilities of LLMs with high data efficiency.
There have been explorations to improve LLMs’ performance in reasoning tasks, such as solving mathematical puzzles~\cite{shao2024deepseekmath,qwen2math,ying2024internlm} and coding~\cite{zhang2024o1code,zhang2024codedpo}.
Furthermore, Visual-RFT~\cite{liu2025visual} pioneers the enhancement of reasoning and visual perception in Large Vision Language Models with limited data.
In the recommendation scenario, compared with SFT, the RL method requires less data to learn a reasoning strategy with good generalization, making it suitable for cold start scenarios and user simulation.
To our knowledge, we are among the first to apply RFT of LLMs for user simulation and video recommendation.

\vspace{-0.3cm}
\section{Method}
\vspace{-0.2cm}
As shown in Fig. \ref{fig:intro}, our VRAgent-R1 framework mainly consists of two components: the Item Perception Agent (IP Agent) for video modeling and the User Simulation Agent (US Agent) for user modeling. In the following sections, we will detail how each agent works and how they collaborate to achieve better user simulation.

\vspace{-0.2cm}
\subsection{Item Perception Agent (IP Agent)} 

\begin{figure}[t] 
    \centering
    \setlength{\abovecaptionskip}{0.1cm}
    \includegraphics[width=\linewidth]{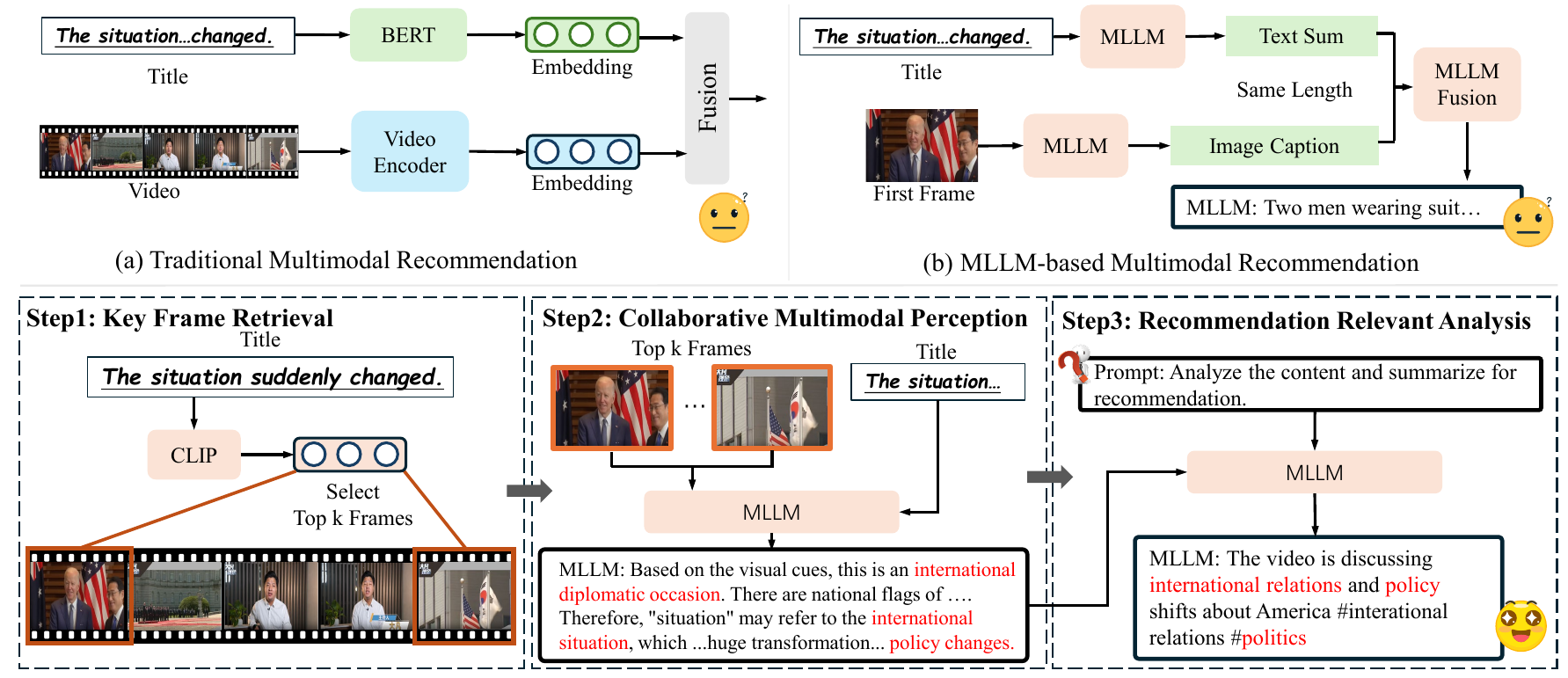}
    \caption{\textbf{Video understanding by the IP Agent.} We simulate the human video comprehension process
   through a progressive approach involving retrieval, collaborative perception, and analysis,
   so as to obtain a summary of the key video information that is applicable for recommendation.
    }
    \vspace{-0.3cm}
    \label{fig:MLLM}
     \vspace{-0.3cm}
\end{figure}

Existing video representation learning methods typically process visual and textual information separately by inputting them into distinct encoders for later feature fusion. 
However, due to the heterogeneity and the imbalance in information volumes between the two modalities, 
it is prone to modality competition, which in turn leads to a suboptimal semantic space for item representations~\cite{zhou2023comprehensive}.
To accurately localize and extract the high-level semantics of video for more precise recommendations, 
we propose a progressive approach that utilizes MLLMs to gradually mine video information through key frame retrieval, collaborative multimodal perception, and recommendation-relevant analysis, as illustrated in Fig. \ref{fig:MLLM}.

\textbf{Key Frame Retrieval (KFR).}
Given that videos contain a wealth of visual information, directly utilizing all video frames would introduce substantial redundant information and result in low computational efficiency. 
To identify the most crucial information in the visual representation while ensuring the algorithm's efficiency, 
we uniformly sample 10 frames from the video. 
Subsequently, we employ CLIP~\cite{clip} to compute the visual-text similarity scores between these sampled frames and the video title. 
The frames with the top 3 highest CLIP scores are then identified as the key visual information for the video's representation.

\textbf{Collaborative Multimodal Perception (CMP).}
After obtaining the retrieved frames,
the next step is to identify the specific events and high-level semantics conveyed in the video. 
Our approach stands out from previous methods 
by fully leveraging the multi-modal understanding capabilities of the MLLM. 
Specifically, since some titles do not directly reflect the video’s topics,
we input both the retrieved frames and titles into the MLLM and prompt it to understand the semantic context implied by the titles.
During this process, MLLM can provide relevant explanations of the title and offer supplementary information. 
For instance, in Fig. \ref{fig:MLLM}, the term "situation" might pertain to international relations, while “change” could imply shifts in national policy. 
It is important to note that neither modality’s embedding could independently capture such nuanced semantics. 
Thus, the MLLM can now clearly comprehend the video information, including the main characters, general events, video genre, and the sentiment expressed in the video.

\textbf{Recommendation Relevant Analysis (RRA).}
The captions initially generated by MLLM may not be well-suited for the specific recommendation scenario, as they may contain excessive redundant explanations or even hallucinations.
To address this issue, we prompt the model to analyze the detailed video content jointly with the characteristics of scenario, as well as focusing on the key information that users are most likely to find interesting.
The model then reformulates the video content into a concise and precise caption limited to approximately 35 words, which is close to the average length of the original titles. This process helps filter out unimportant details, resulting in a unified and comprehensive video caption. For more details, please refer to Appendix A.1. 
Note that the reformulated video caption not only can be used to enhance the representation learning for items, but also assist the process of user behavior simulation.

\vspace{-0.1cm}
\subsection{User Simulation Agent (US Agent)} 
\vspace{-0.2cm}

After modeling the video items, we then consider simulating human behavior to refine the recommendation results.
Although numerous LLM-based user agents are available~\cite{zhang2025llm,zhang2024agent4rec,wang2025user,xiang2024simuser},
most of them simply prompt frozen LLMs to generate fixed user profiles and make predictions without incorporating downstream feedback.
Therefore, LLMs can not be optimized and the simulation outcomes may be unrealistic and prone to hallucinations.
Moreover, simple supervised fine-tuning only enables models to memorize answers.
Due to the lack of in-depth analysis of user behavior, this approach yields limited accuracy and lacks interpretability.
To better align the model with the user decision-making process, we innovatively employ reinforcement learning to fine-tune the LLM within a simulated recommendation environment.

\textbf{Environment and User Modeling.}
First, we identify the modeling of the recommendation environment and personalized user 
as shown in Fig .\ref{fig:intro}.
The environment aims to simulate a realistic recommendation scenario by generating candidate videos for the user, while the US Agent simulates the user to perform specific tasks.
More specifically,
for a user with $N$ behaviors (including watched videos and corresponding comments), the first $N$-1 behaviors are used for user profile modeling, and the $N$-th behavior is regarded as the prediction target.
We use SASRec~\cite{ni2023content} 
to recommend 10 video items based on the user's historical behaviors, simulating a rough recall process.
Then \textit{m} items are randomly selected as negative samples, and the real $N$-th item of user behavior is regarded as the positive one.
These \textit{m}+1 items collectively form the candidate video list for future tasks, which we will discuss in the following section. 
However, processing multiple video and text sequences simultaneously is challenging for the MLLM. 
To address this issue, the IP Agent converts the relevant multimodal videos into a textual format, enabling the US Agent to process the long text sequence instead.
For a given task in the RFT process, we prompt the US Agent to first thoroughly analyze the user behavior to formulate a unique user status \(s\)  (e.g., preferences and emotions) through CoT reasoning. 
The US Agent then analyzes the candidate videos and performs the appropriate action. 
Compared to previous methods, the user profile modeling here is dynamically updated based on task rewards, which allows for a learnable and more accurate simulation.

\textbf{Task and Reward.}
%
%
To align the US Agent with real user preferences,
we design two specific tasks for RFT, i.e., User Preference Judgment and Next Video Selection.
In the first task, following settings of previous methods~\cite{MLLM-MSR,zhang2024agent4rec,zhang2025llm}, 
the agent is given an item from the candidate list and then prompted to judge whether the user would like the recommended video.
The action space \(\mathcal{A}\) consists of {"Yes" and "No"}, corresponding to positive items and negative items, respectively.
The Reward $R_1$ for this task comprises two parts: the format reward $R_{\text{format}}$ and the judgment reward $R_{\text{jud}}$,
\begin{equation}    
R_1 = R_{\text{format}} + R_{\text{jud}}.
\end{equation}
The format reward ensures the model adheres to the required response format,
i.e.,
\textit{
<think>the CoT thinking process</think>,
<answer>the final answer</answer>}.
If the model's answer is correctly formatted, it receives a score of 1; otherwise, scores are assigned according to the format specification presented in Appendix A.2. 
Additionally, we use a post-processing function \textit{f} to parse the answer within the \textit{<answer>} tag into a legal action,
and check if it matches the ground truth. Here, $ R_{jud}$ will be 1 for a correct simulation and -1 for a wrong situation.

In the second task,
the agent first reviews all candidate items and selects the video that the user is most likely to watch next via prompt engineering.
The action space \(\mathcal{A}\) consists of choosing one item from the \textit{m+}1 candidate videos.
The reward $R_2$ for this task includes the format reward $R_{\text{format}}$ and 
the selection reward  $R_{\text{sel}}$, 
\begin{equation}
    R_2 = R_{\text{format}} + R_{\text{sel}}.
\end{equation}
Given the larger action space of $R_2$  compared to $R_1$,
this task is more complex, and we assign a score of 2 for correctly selecting the positive video to provide a higher reward.


\textbf{GRPO Training.}
%
We employ Group Relative Policy Optimization (GRPO)~\cite{shao2024deepseekmath} framework to train the agent, which compares groups of candidate responses directly, without requiring a critic model to evaluate policy performance.
Given a problem \(q\) for the model \(\pi_{\theta}\), it samples to generate a group of distinct answers \( {o_i} \), where \(i = 1, 2, \dots, G\) and \(G\) is the sampled number in the group.
Each answer involves different CoT reasoning for the user status and final answer, and we compute the corresponding reward \( {r_i} \).
By comparing the relative advantage of the \textit{i}-th answer $ \hat{A}_{i}$,
\begin{equation}
  \hat{A}_{i}=\frac{r_i - \text{mean}(\textbf{r})}{\text{std}(\textbf{r})} 
\end{equation}
\(\textbf{r} = \{r_1, r_2, \dots, r_G\}\), 
GRPO encourages the model to select the answer with higher reward within the group (more details are in Appendix A.4).
We initially train the agent with an easy judgment task,
then introduce the selection task.
Via such a progressive training manner,
the agent learns from simpler to more complex tasks, and the CoT process is gradually optimized,
which provides thoughtful and interpretable recommendations for the user behavior.

\vspace{-2mm}
\section{Experiments}
\vspace{-2mm}
\textbf{Datasets and Metrics.}
We have conducted extensive experiments on MicroLens-100K~\cite{ni2023content}, an open-source, real-world video recommendation dataset that includes 100,000 users, 19,738 items, and 719,405 interactions, with a sparsity level of 99.96\%. 
Notably, MicroLens is the first micro-video recommendation dataset to provide original video content, which enables us to analyze videos from raw frames. The average video duration is 161 seconds, featuring rich visual content.
For traditional recommendation metrics across the entire dataset, we report standard recommendation metrics such as HR@10, NDCG@10, HR@20, and NDCG@20. 
For the evaluation of user behavior simulation, we report binary classification metrics such as accuracy (Acc) and F1 Score for preference judgment, as well as selection accuracy for next video selection.

\textbf{Implementation Details.}
We employ Qwen2.5-7b~\cite{yang2024qwen2} for both the IP Agent and the US Agent.
For video recommendation, we follow the official code of MicroLens-100K\cite{ni2023content} benchmark for evaluation,
and select the frames with the top 3 similarity scores generated by the IP Agent.
In user simulation,
we designate the actual last item of user behavior as the positive sample and employ SASRec~\cite{ni2023content} to generate the top 10 recommended items, from which \textit{m} items are randomly chosen as negative samples to more accurately mimic real-world recommendation scenarios.
We deploy 4 A100 80G GPUs for the reinforcement fine-tuning of the US Agent using information from 2000 users.
The training configuration includes a batch size of 16, 16 sampled policy rollouts, and a KL coefficient of 0.001. 
We set the maximum number of input and response tokens to 2048 and train the model for 4 epochs, which takes approximately 10 hours. 
User simulation evaluation is conducted on 1000 randomly selected cold-start users and repeated three times to report the average performance.
The specific prompts used for instructing models are detailed in Appendix A.1.

\begin{table*}[t]
\caption{\textbf{Comparison results on MicroLens-100K.} The fusion of different modality features is achieved by weighted pooling. The underline \textit{T}, \textit{I}, \textit{V}, \textit{F} correspond to text, image, video, and fusion features, respectively. Our MLLM-enhanced features successfully boost the performance of different traditional sequence recommendation methods.}
\small
\centering
\resizebox{\linewidth}{!}{
\begin{tabular}{l|l|cccc} 
\toprule
Class & Model & HR@10 & NDCG@10 & HR@20 & NDCG@20 \\
\midrule
\multirow{3}{*}{\textbf{IDRec(CF)}} & DSSM~\cite{ddsm} & 0.0394 & 0.0193 & 0.0654 & 0.0258 \\
& LightGCN~\cite{he2020lightgcn} & 0.0372 & 0.0177 & 0.0618 & 0.0239 \\
& DeepFM~\cite{guo2017deepfm} & 0.0350 & 0.0170 & 0.0571 & 0.0225 \\
\midrule
\multirow{3}{*}{\textbf{IDRec(SR)}} & NextItNet~\cite{nextitnet} & 0.0805 & 0.0442 & 0.1175 & 0.0535 \\
& GRU4Rec~\cite{hidasi2015session} & 0.0782 & 0.0432 & 0.1147 & 0.0515 \\
& SASRec~\cite{kang2018self} & 0.0909 & \underline{0.0517} & 0.1278 & 0.0610 \\
\midrule
\multirow{5}{*}{\textbf{Modality}} & NextItNet$_{V}$~\cite{ni2023content} & 0.0862 & 0.0466 & 0.1246 & 0.0562 \\
& SASRec$_{T}$~\cite{ni2023content} & 0.0916 & 0.0490 & 0.1343 & 0.0598 \\
& SASRec$_{I}$~\cite{ni2023content} & 0.0942 & 0.0511 & 0.1358 & 0.0613 \\
& SASRec$_{V}$~\cite{ni2023content} & 0.0948 & 0.0515 & \underline{0.1364} & 0.0619 \\
& SASRec$_{F}$~\cite{ni2023content} & \underline{0.0953} & \underline{0.0517} & 0.1362 & \underline{0.0623} \\
\midrule
\multirow{3}{*}{\textbf{MLLM Enhanced}} & SASRec~\cite{kang2018self}+MLLM-MSR~\cite{MLLM-MSR} & 0.0606 & 0.0351 & 0.0911 & 0.0446 \\
& \textbf{NextItNet~\cite{nextitnet}+Ours} & 0.0884 & 0.0478 & 0.1278 & 0.0583 \\
& \textbf{SASRec~\cite{kang2018self}+Ours} & \textbf{0.0994}$_{\textcolor{red}{+4.30\%}}$ & \textbf{0.0548}$_{\textcolor{red}{+6.00\%}}$ & \textbf{0.1418}$_{\textcolor{red}{+3.96\%}}$ & \textbf{0.0655}$_{\textcolor{red}{+5.14\%}}$ \\
\bottomrule
\end{tabular}
}
\vspace{-0.2cm}
\label{tab5}      
\end{table*}

\begin{figure}[!tbp]
    \centering
    \includegraphics[width=0.9\linewidth]{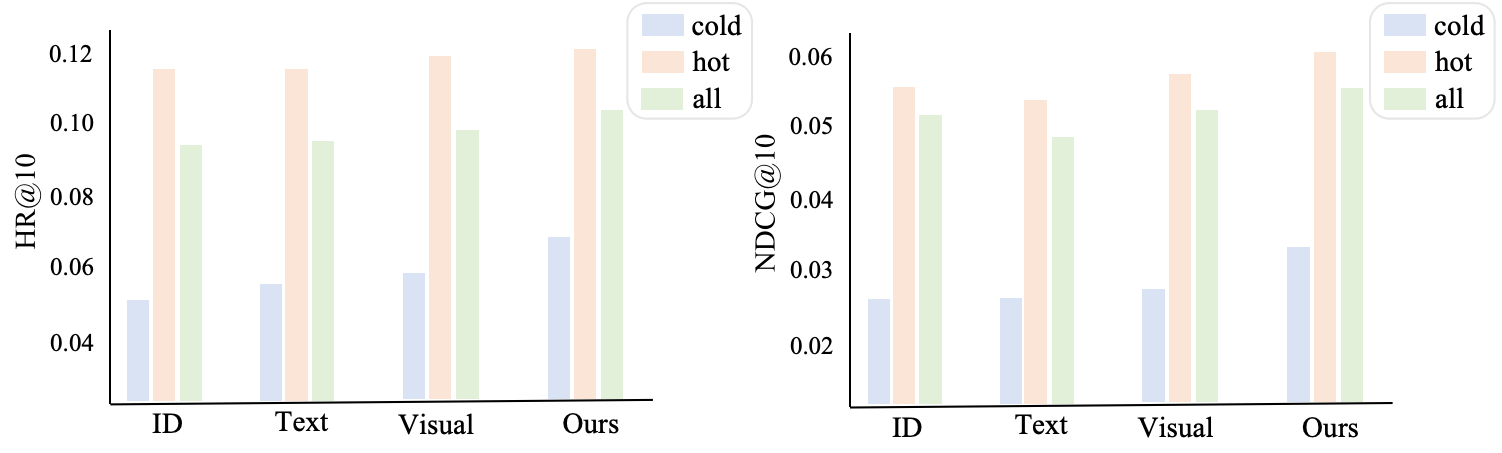}
    \vspace{-0.2cm}
    \caption{\textbf{User Group Performance.} Our method has a great advantage in modeling cold-start users due to the good multimodal understanding, outperforming the baseline by more than 10\%. }
    \label{fig:cold}
    \vspace{-0.4cm}
\end{figure}

\subsection{Performance Comparison for Video Recommendation}
In this section, we evaluate the effectiveness of our IP Agent on the video recommendation baseline, following the experimental protocols in MicroLens~\cite{ni2023content}.
We use the MLLM to generate outputs for collaborative reasoning and understanding, 
thereby enhancing the content of the original titles.
As shown in Table \ref{tab5}, sequential recommendation models that incorporate multimodal information outperform both Collaborative Filtering (CF) models and simple ID embedding-based models.
However, traditional multimodal methods rely on late fusion techniques (e.g., addition or concatenation) of ID, textual, and visual embeddings. 
Due to the significant differences between modalities, such coarse-grained fusion fails to fully leverage multimodal information. 
Moreover, the lack of in-depth understanding of video frames means that these fusion methods only achieve similar performance to single-modality approaches.
We also experiment with directly using the MLLM to generate a detailed caption for the cover image, as in MLLM-MSR~\cite{MLLM-MSR}. However, these captions often focus on unimportant details in the image, failing to capture key information and leading to a significant decline in performance.
In contrast, our IP Agent integrates the pre-trained world knowledge of the MLLM to collaboratively understand multimodal content. It summarizes key information in a brief and unified format, which helps to boost HR@10 and NDCG@10 by 4.3\% and 6.0\%, respectively, compared to previous state-of-the-art methods.




Considering that our method establishes a comprehensive understanding of multimodal items, it should also be helpful in handling cold-start users. Accordingly, we evaluate the performance of different methods on cold-start users (i.e., users with no more than 5 interacted videos), as shown in Fig. \ref{fig:cold}.
It is evident that the ID-based method suffers the most significant performance degradation for cold-start users, as it relies solely on ID embeddings without incorporating multimodal information. In contrast, our method demonstrates a clear advantage in modeling cold-start users, outperforming the baseline by more than 10\%.

\begin{table}[t] 
\centering
\caption{\textbf{Evaluation on User Simulation.} Our RFT method outperforms previous prompt and SFT-based simulation, with less training data and higher accuracy. \textit{m} is the number of negative samples, and SFT methods have higher Recall scores since they tend to give positive answers. }
\footnotesize
    \centering
    \small
    \begin{tabular}{l|cccc|cc}
    
        \toprule
        Method & Acc & Recall &  Pre & F1  &  Acc$_{m=3}$  & Acc$_{m=4}$   \\
        \midrule
        
        GPT-4o~\cite{gpt4o} &  0.535   &   0.480    & 0.533      & 0.505 & 0.307 & 0.269                                   \\
        DeepSeek-R1~\cite{guo2025deepseek} &   0.528  & 0.556      &  0.526     &  0.541 & 0.264 & 0.231        \\
       Qwen2.5-7b~\cite{yang2024qwen2}  & 0.491 & 0.512  & 0.490 &  0.500  &   0.245  &  0.197\\
         LLM\_Simulator ~\cite{zhang2025llm}  & 0.523   &  0.539  & 0.519  & 0.529 & - & -  \\
         Agent4Rec~\cite{zhang2024agent4rec}  &  0.528  & 0.482  &  0.52    & 0.501  & - & -  \\
        TALLREC (SFT)~\cite{bao2023tallrec}  &  0.537    &  \textbf{0.913} & 0.521   & 0.663  &   -  &  -  \\
        MLLM-MSR (SFT)~\cite{MLLM-MSR}  & 0.585  &  0.882  & 0.553   &   0.679   & 0.442 & 0.381   \\

        \rowcolor{gray!20}
        \textbf{VRAgent-R1}   &  \textbf{0.715} & 0.760 & \textbf{0.697} & \textbf{0.727} &  \textbf{0.641}  & \textbf{0.602} \\
        \bottomrule
    \end{tabular}
    \vspace{-0.5em}
    \label{tab:simulation}      
\end{table}

\begin{table*}[!tbp]
\renewcommand{\arraystretch}{1.2}
\centering
\begin{minipage}[t]{0.47\textwidth}

\caption{\textbf{Preference evaluation comparison on MovieLens with our US Agents.}}
\vspace{-0.5em}

\centering
\resizebox{\textwidth}{!}{ 
\begin{tabular}{l|cccc}
\toprule

Method &  Acc & Recall &  Pre & F1  \\\midrule
GPT-4o~\cite{gpt4o}  &  0.584   &   0.626   & 0.577     & 0.600 \\
RecAgent~\cite{wang2025user} & 0.581 & 0.639 & 0.604 &  0.621 \\
 Agent4Rec~\cite{zhang2024agent4rec} &  0.691 &  0.746 & 0.691 &  0.698  \\
\rowcolor{gray!20}
\textbf{VRAgent}  &  \textbf{0.832} & \textbf{0.846} & \textbf{0.823} & \textbf{0.834}\\
 
\bottomrule
\end{tabular}
}
\vspace{-1.0em}
\label{tab:movielens}         
 \end{minipage}
\begin{minipage}[t]{0.52\textwidth}
\caption{\textbf{Optimizing RSs with VRAgent-R1.} 
\vspace{-0.1em}
}
\centering
\resizebox{\textwidth}{!}{ 
\begin{tabular}{l|cc|cc} 
\toprule
\multirow{2}{*}{Method}  & \multicolumn{2}{c|}{\textbf{All}} & \multicolumn{2}{c}{\textbf{Cold}}  \\
& HR@10 & NDCG@10 & HR@10 & NDCG@10  \\
\midrule
Original & 0.0916 & 0.0490 & 0.0586 & 0.0278 \\
+IP Agent & 0.0994 & 0.0548 & 0.0663  & 0.0318  \\
\midrule

+US dislike &  0.0988  & 0.0543  & 0.0654   & 0.0311   \\
\rowcolor{gray!20}

\textbf{+US like } & \textbf{0.1003} & \textbf{0.0554} &  \textbf{0.0678} & \textbf{0.0330}  \\
\bottomrule

\end{tabular}
}
\label{tab:feedback}     
\end{minipage}
 \vspace{-1em}
\end{table*}

\vspace{-0.15cm}
\subsection{Evaluation on User Simulation}

In this subsection, we assess the performance of VRAgent-R1 and other user simulators in accurately modeling user behavior through two tasks: user preference prediction (estimating whether a user will like recommended items) and next video selection (choosing the most likely video a user will click next from candidates). Tab. \ref{tab:simulation} presents the results comparing our method with commercial models (GPT-4o~\cite{gpt4o}, DeepSeek-R1~\cite{guo2025deepseek}) and several user simulation baseline methods.
As shown in the table, directly prompting frozen LLMs to summarize user preferences based on textual inputs and then predict the answer yields poor results. This is because these LLMs rely solely on limited information from video titles and suffer from serious hallucination problems, leading to performance only slightly better than random guessing.
For the SFT methods, such as ALLREC~\cite{bao2023tallrec} and MLLM-MSR~\cite{MLLM-MSR}, training the models to predict user behavior based on summarized user preferences shows improved performance after fine-tuning with an additional 20,000 users' information. 
MLLM-MSR performs better due to the inclusion of extra cover image information, but the results are still unsatisfactory, and the final model can only provide binary answers ("Yes" or "No") without a reasoning process.
On the contrary, our VRAgent-R1 method achieves significantly higher prediction accuracy, e.g., about 45\% higher for the next video selection than MLLM-MSR, despite using less training data. 
Additionally, our model can be applied to different reasoning tasks without losing generality.

To verify the effectiveness of our method across different domains, we conduct user preference simulation tests~\cite{zhang2024agent4rec} on the widely used MovieLens-1M~\cite{harper2015movielens} dataset. 
Since the understanding of movie content in this dataset primarily relies on text descriptions, we evaluate the performance using only the US Agent, without involving the IP Agent for multimodal processing. 
The results in Tab. \ref{tab:movielens} indicate that in text-dominated movie recommendation scenarios, our approach also significantly outperforms previous simulation agents~\cite{wang2025user,zhang2024agent4rec}.
More discussion could be referred in Appendix A.3.

\vspace{-0.3cm}
\subsection{Optimizing RSs with VRAgent-R1}
\vspace{-0.2cm}
We conduct a preliminary experiment to assess whether VRAgent-R1's simulation could enhance recommendation systems (RSs).
Specifically, we randomly select 8,000 cold-start users and provide VRAgent-R1 with the top 10 items recommended by the original RSs. VRAgent-R1 simulates user decisions on which videos they might watch and which they would dislike. 
This simulated behaviors are then used to supplement the modeling of cold-start users and update the RSs.
As shown in Tab. \ref{tab:feedback}, incorporating feedback on user-liked videos improves recommendation performance. In contrast, simulated interactions with user-disliked videos have a negative impact. 
This outcome effectively demonstrates the feedback-driven recommendation augmentation process.


\vspace{-0.1cm}
\textbf{Ablation on the IP Agent.}
In this part, we ablate the progressive steps taken in our IP Agent.
Specifically, "w/o KFR" denotes the absence of the key frame retrieval process, where we replace it with three randomly selected adjacent frames.
"w/o CMP" indicates the removal of collaborative multimodal perception. In this scenario, 
the MLLM is employed to analyze visual frames and titles independently. 
"w/o RRA" signifies the exclusion of recommendation-relevant analysis, leading to the direct utilization of detailed long textual outputs for video comprehension.
As shown in Tab. \ref{tab:cot},
both key frame extraction and collaborative multimodal interaction significantly enhance video modeling. The analysis process is also crucial, as excessive unimportant details can be noisy and degrade recommendation performance.

\textbf{Ablation on the US Agent.}
We then conduct ablation studies to assess the impact of each component of VRAgent-R1 on simulation performance, as detailed in Tab. \ref{tab:ablation}.
First, we remove the CoT reasoning from the RFT process,
and then the LLM directly predicts answers based on the user's historical sequence, bypassing analysis of video contents and user status. This leads to a significant performance drop, underscoring the importance of CoT in reasoning tasks.
Next, we evaluate the impact of user comments which aid in more accurate user modeling.
Results show that the absence of them causes roughly 4\% performance decline.
Finally, we ablate the IP Agent, which is responsible for multimodal understanding.
We can observe that the IP Agent boosts performance by approximately 6\% over the baseline relying on original textual information.

\begin{table*}[t]
\renewcommand{\arraystretch}{1.2}
\centering
\begin{minipage}[t]{0.49\textwidth}
\vspace{-0.65em}
\caption{\textbf{Ablation on the IP Agent.}}
\vspace{-0.7em}
\centering
\Large
\resizebox{1\textwidth}{!}
{ 
\begin{tabular}{l|cccc} 
\toprule
Method & HR@10 & NDCG@10 & HR@20 & NDCG@20  \\
\midrule
Baseline & 0.0916 & 0.0490 & 0.1343 & 0.0598  \\ \midrule
w/o KFR & 0.0980 & 0.0542  & 0.1422  & 0.0646   \\   
w/o CMP &  0.0960 & 0.0519 & 0.1398 & 0.0629  \\
w/o RRA & 0.0816 & 0.0466 & 0.1182 & 0.0523    \\
\rowcolor{gray!20}
\textbf{VRAgent-R1} & \textbf{0.0994} & \textbf{0.0548} & \textbf{0.1448} & \textbf{0.0655}   \\
\bottomrule
\end{tabular}
}
\label{tab:cot}      
 \end{minipage}
\begin{minipage}[t]{0.45\textwidth}
\vspace{-1.2em}
\caption{\textbf{Ablation on the US Agent.}}
\vspace{-0.5em}

    \Large
     \resizebox{0.95\textwidth}{!}{ 
    \begin{tabular}{l|cc|cc}
        \toprule
        Method  & Acc & F1  &  Acc$_{m=3}$  & Acc$_{m=4}$   \\
        \midrule
        Baseline & 0.491 &  0.500 & 0.245 & 0.197 \\
        SFT &  0.585 & 0.679 & 0.442 & 0.381   \\ \midrule
        
       w/o CoT Reasoning & 0.580 & 0.592     & 0.324  & 0.263   \\
      w/o Comment & 0.695 &    0.705     & 0.618  & 0.577  \\
        w/o IP Agent & 0.680    &  0.686   & 0.609  &  0.569   \\
        \rowcolor{gray!20}
         \textbf{VRAgent-R1}  & \textbf{0.715} &    \textbf{0.727}    & \textbf{0.646}  & \textbf{0.602}    \\
        \bottomrule
    \end{tabular}
    }
     \label{tab:ablation}      
\end{minipage}
\vspace{-0.5cm}
\end{table*}

\vspace{-0.3cm}
\section{Conclusion}
\vspace{-0.2cm}
%
In this paper, we propose a novel VRAgent-R1 framework for user simulation in video recommendation.
It first utilizes an MLLM to collaboratively understand the retrieved multimodal content with pre-trained world knowledge,
then analyzes the history sequence to establish user status and make final decision through RFT.
By exploring different strategies and using real user decisions as verifiable rewards under different tasks, our VRAgent-R1 method achieves significant improvements in user behavior simulation for video recommendation.
It outperforms SFT with minimal data and shows strong generalization. 
As a pioneering work, this study demonstrates the potential of applying RFT to LLMs in recommendation systems.

\vspace{-0.1cm}
\textbf{Limitations and Future Directions.}
%
Although the dataset used in this paper focuses on video recommendations, our VRAgent-R1 paradigm can be easily extended to various domains such as games, news, and e-commerce. 
Currently, the user action space in our experiments is relatively limited due to the dataset properties.
In future work, we plan to explore adding more user behaviors, such as clicks and retention, to achieve more realistic user-system interactions. 
Moreover, verifying whether better user simulation feedback can further optimize existing recommendation systems is also a crucial direction for future exploration.

\small
\bibliographystyle{IEEEtran}
\bibliography{ref.bib}

\newpage
\appendix

\section{Technical Appendices and Supplementary Material}
Technical appendices with additional results, figures, graphs and proofs may be submitted with the paper submission before the full submission deadline (see above), or as a separate PDF in the ZIP file below before the supplementary material deadline. There is no page limit for the technical appendices.

\subsection{Prompt Templates for the Agents}

In this section, we present the comprehensive set of prompt templates utilized in the experiments across various stages, as shown in Tab.~\ref{tab:all-prompts}. These prompts are meticulously designed to guide the Agent in effectively performing its tasks, ensuring seamless interactions and accurate responses.

\begin{small}
\captionsetup{width=\textwidth}
\begin{longtable}{p{3.8cm}|p{8cm}}
\toprule
\textbf{Prompt} & \textbf{Content} \\ \midrule
\endhead
Prompt to decide which parts of the video to focus on & 
You are a helpful assistant to help to understand a video.
These are key \textbf{\{frames\}} from a video, and the title of the video is: \textbf{\{title\}}. Pay special attention to content related to the title. 

\\ \midrule
Prompt for collaborative \newline perception & 
Based on the textual and visual contents, 
identify what visual content aligns with or extends beyond the title's description, note any discrepancies or additional context provided by the visuals.
Now combined with your knowledge and understanding,
give the key information about the video, including: main characters, core event and emotional appeal.
 
  \\ \midrule

Prompt to generate summarization for recommendation & 
If you want to recommend the video, 
create a concise summary within 35 words on what the viewer would be interested in,
following this format: [Core Content Description]  + [Refined Tags].
The content should be clear and elegant, 
and tags should be brief and accurate to reflect the video topic.
Here is a good example:
"the girl just drove the wrong way, but did not expect to encounter terrible things \# thriller movie \# movie commentary"

\\ \midrule
Prompt for User Simulation &
You are a helpful assistant. The assistant first thinks about the user's video watching history and the comments, analyzes their current status, such as preferences and purpose, and predicts:
which video they are most likely to watch next from the given candidates / if they like the next video. The reasoning process and answer are enclosed within <think> </think> and <answer> </answer> tags, respectively, i.e., <think> reasoning process here </think><answer> answer here </answer>. 
After thinking, when you finally reach a conclusion, give the user status and the answer you predict within <answer> </answer> tags. i.e., <think> (1) User\_status:... </think><answer>(2) Next\_video:...  </answer>.
    \newline
    User's viewing history: \textbf{\{history\_str\}}
    \newline
    Candidate videos for the next watch: \textbf{\{candidates\_str\}}

\\
\bottomrule

\caption{The prompt templates used in our VRAgent-R1.}
\label{tab:all-prompts} 
\end{longtable}
\end{small}

\subsection{Reward Score Computation}
In this section, we give the concrete computation for the three reward scores as follows:
\begin{equation}
    R_{format} = 
\begin{cases}
1, & \text{if the answer follows the standard format} \\
0.5, & \text{if the order of <think>,<answer> tags is wrong} \\
0, & \text{if missing <think> or <answer> tags} \\
-1, & \text{if the answer cannot be parsed or is missing}
\end{cases}
\end{equation}

\begin{equation}
    R_{jud} = 
\begin{cases}
1, & \text{if the agent preference matches the ground truth} \\
-1, & \text{if the agent preference mismatches the ground truth} \\
-1, & \text{if the agent preference cannot be parsed or is missing}
\end{cases}
\end{equation}

\begin{equation}
    R_{sel} = 
\begin{cases}
2, & \text{if the agent selection matches the ground truth} \\
-1.5, & \text{if the agent selection mismatches the ground truth} \\
-2, & \text{if the agent selection cannot be parsed or is missing}
\end{cases}
\end{equation}

\begin{figure}[h]
    \centering
    \setlength{\abovecaptionskip}{0.1cm}
    \includegraphics[width=\linewidth]{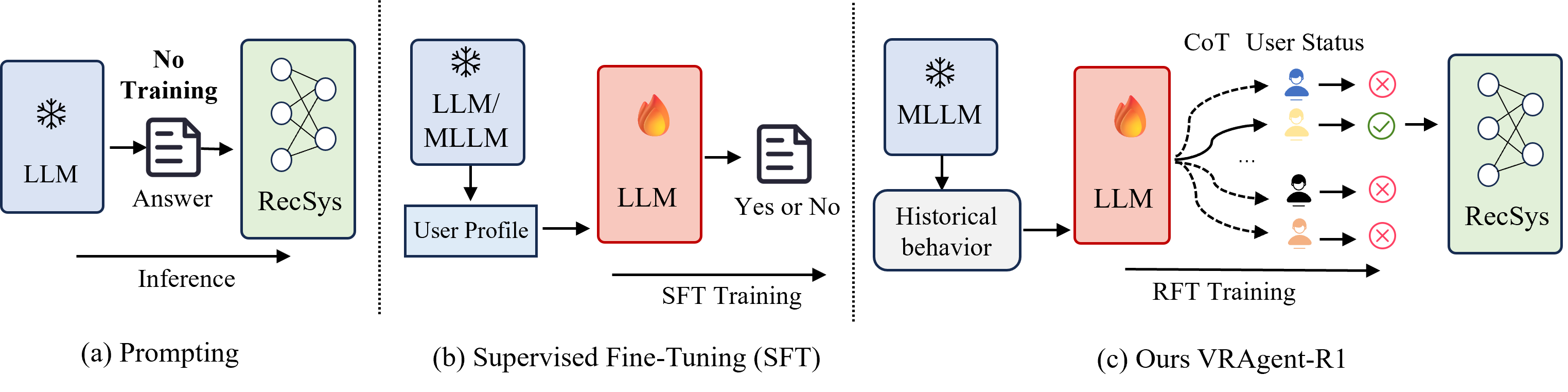}
    \caption{\textbf{Comparison with other user simulation for recommendation.} }
    \vspace{-0.3cm}
    \label{fig:RFT}
    \vspace{-0.2cm}
\end{figure}
\subsection{Additional Experiment Discussion}

\paragraph{Experiments on MovieLens.}
To verify the generality of our method in other domains, we also conduct user preference simulation tests on the widely used MovieLens-1M~\cite{harper2015movielens} dataset,
which contains 1 million ratings from 6000 users on 4000 movies, 
and ratings above 3 are considered a positive \textit{like} signal.
Note that, though this dataset is related to movies, 
the understanding of movie contents mainly relies on text descriptions,
while the visual information is not that important. 
Therefore, in this experiment, we only use the US Agent to do the evaluation, without considering the IP Agent for multimodal processing.
We follow the setting of Agent4Rec~\cite{zhang2024agent4rec}, 1000 simulated users are randomly assigned 20 items, with varying ratios \textit{1:m} of positive and negative items.
In our main paper, the reported results are all in a \textit{1:1} ratio, here we also report the performance in \textit{1:3} setting.

\begin{table}[h] 
\centering
\caption{\textbf{Preference evaluation comparison on MovieLens with our US Agents.} }
\footnotesize
    \centering
    \begin{tabular}{l|cccc|cccc}
    
        \toprule
        
       \multirow{2}{*}{\textbf{Method}} &   \multicolumn{4}{c|}{1:1}  &   \multicolumn{4}{c}{1:3} \\
       & Acc & Recall &  Pre & F1 & Acc & Recall &  Pre & F1  \\
        \midrule
        GPT-4o~\cite{gpt4o}  &  0.584   &   0.626   & 0.577   & 0.600 & 0.523 & 0.701 & 0.308 & 0.428
        \\ 

        RecAgent~\cite{wang2025user} & 0.581 & 0.639 & 0.604 &  0.621 &  0.508 & 0.740 & 0.399 & 0.518
        \\
         Agent4Rec~\cite{zhang2024agent4rec} &  0.691 &  0.746 & 0.691 &  0.698 &  0.668 & 0.762 &0.421 & 0.543  
         \\

        \textbf{Ours}  &  \textbf{0.832} & \textbf{0.846} & \textbf{0.823} & \textbf{0.834} &  \textbf{0.806} &   \textbf{0.821}    &  \textbf{0.579}      &    \textbf{0.679}
        \\
       
        \bottomrule
    \end{tabular}
    \vspace{-1.0em}
    \label{tab:movielens_expand}      
\end{table}

We can learn from the Tab.~\ref{tab:movielens_expand} that the user simulation results on MovieLens are better than those on MicroLens, indicating that the video recommendation task on MicroLens is more complex and challenging, where the multimodal information should be considered.
And for textual dominated movie recommendation, our approach still has a great advantage compared with previous prompt-based simulation Agents~\cite{wang2025user,zhang2024agent4rec} like in Fig. \ref{fig:RFT},
the prompt-based agents could not be optimized accordingly to recommendation, while our VRAgent-R1 learns to think deeply to simulate more realistic user behavior.


\paragraph{Statistical Significance}
For the video recommendation on MicroLens-100k, we follow the official implementation~\cite{ni2023content}, 
with the IP Agent enhanced video understanding, we train the SASRec~\cite{kang2018self} with text information for three times, every time the test performance shows almost the same results, 
demonstrating the reproducibility of our experiments.
For the user simulation, we use three random seeds to select 1000 cold-start users for evaluation,
with 1-sigma error bars of 0.002, 0.005 for Acc and F1 in user performance judgment, 
0.003 and 0.003 for Acc$_{m=3}$ and Acc$_{m=4}$ in next video selection.
And the results consistently show that the performance of our method significantly outperforms previous agents (statistical tests using paired t-tests with 95\% confidence intervals yield a p<0.002). 


\subsection{Reinforcement Learning in Recommendation}
\paragraph{Previous Works.}
In this section, we give a brief review of previous reinforcement learning methods for recommendation systems.
Before the prevalence of LLMs, there were already many methods involving reinforcement learning (RL) for various objectives in recommendation.
For example,
previous methods use GANs~\cite{shi2019virtual,chen2019generative} or transformers~\cite{zhao2023kuaisim} to simulate the user behavior,
while they still involve the fitting of features essentially, relying on a large amount of data for reinforcement learning.
They are limited to predefined tasks, unable to simulate users' emotions and thinking processes, nor can they provide interpretive feedback.
Recently, with the strong logical reasoning ability of LLMs, many researchers have begun to explore how to integrate RL and LLM to improve recommendation systems.
$P^{4}$LM~\cite{jeong2023factual} applies RLHF to align a language model with factuality, personalization, and appeal to provide more interpretive explanations in movie recommendation scenario.
RLRF4Rec~\cite{sun2024rlrf4rec} aims to enable the LLMs with the knowledge augmentation recommendation,
and Rec-R1~\cite{lin2025rec} directly bridges the recommendation and LLMs,
by optimizing the LLM with real time reward signals from the recommendation system.
Different from above methods that generate better text outputs to help recommendation, 
we aim to simulate more realistic user behavior in multimodal situations, and our method has more application potential since it's not limited to text-guided recommendation, 
it can not only evaluate the recommendation quality but also improve recommendation results through simulated feedback.

\paragraph{Basic RL Standards for LLM.}
Without loss of generality, 
we adhere to the standard notations presented in the classic works of reinforcement learning~\cite{sutton1998reinforcement,agarwal2019reinforcement}.
More specifically, we use  
\(s \in \mathcal{S}\) to denote the state space, 
\(a \in \mathcal{A}\) to denote the action space,
\(r_k\) to denote the reward function in step k,
\(\mathcal{P}\) to denote the transition dynamics,
\(\pi(a|s)\) is the probability of performing action \(a\) in state \(s\) under policy \(\pi\),
and \(\gamma \in [0,1]\) is the discount factor.
The goal is to maximize the discounted cumulative returns for each trajectory as below,
\begin{equation}
G_t = \sum_{k=t+1}^{T} \gamma^{k-t}r_k
\end{equation}
where T is the maximum step numbers per episode.
Instead of using the classic PPO~\cite{schulman2017ppo} algorithm that requires a critic model to evaluate policy performance,
we use the GRPO~\cite{shao2024deepseekmath} to compare groups of candidate responses directly.
\begin{equation}
\begin{split}
\mathcal{J}_{\text{GRPO}}(\theta)
&= \mathbb{E}_{[q\sim P(Q), \{o_i\}_{i = 1}^{G}\sim \pi_{\theta_{\text{old}}}(O\mid q)]}\\
&\frac{1}{G}\sum_{i = 1}^{G}\frac{1}{|o_i|}\sum_{t = 1}^{|o_i|}
\left\{
\min\left[
\frac{\pi_{\theta}^{i,t}}{\pi_{\theta_{\text{old}}}^{i,t}}\hat{A}_{i,t}, 
\text{clip}\left(\frac{\pi_{\theta}^{i,t}}{\pi_{\theta_{\text{old}}}^{i,t}}, 1 - \epsilon, 1 + \epsilon\right)\hat{A}_{i,t}
\right]
- \beta\mathbb{D}_{\text{KL}}[\pi_{\theta}\|\pi_{\text{ref}}]
\right\}
\end{split}
\end{equation}
\begin{equation}
    \hat{A}_{i,t}=\frac{r_i - \text{mean}(\textbf{r})}{\text{std}(\textbf{r})}
\end{equation}
Given a problem \(q\) for the model \(\pi_{\theta}\), it samples to generate a group of distinct answers \( {o_i} \), where \(i = 1, 2, \dots, G\), \(G\) is the sampled number in the group. 
Each answer has a different length \(|o_i|\).
\({ \pi_{\theta}^{i,t}} \) is the policy probability of decoding the \(t\)-th token of the sampled answer.
The KL term constrains that the distribution of \(\pi_{\theta}\) should not deviate too much from the original policy \(\pi_{\text{ref}}\) by penalty coefficient \(\beta \). 
Here, an optimized KL term is adopted, which has the characteristics of being unbiased and having a small variance.
The clip strategy restricts the ratio between $\frac{\pi_{\theta}}{\pi_{\theta_{old}}}$, and by limiting the ratio within the interval $\varepsilon$, it prevents the new strategy from having large numerical updates. 
\(\textbf{r} = \{r_1, r_2, \dots, r_G\}\), and \(\hat{A}_{i,t}\) is the relative advantage of the \textit{i}-th answer.
Through the optimization of \(\mathcal{J}_{\text{GRPO}}(\theta)\),
GRPO encourages the model to choose the answer with higher reward within the group.

\subsection{Additional Visualization}

\begin{figure}[t]
    \centering
    \setlength{\abovecaptionskip}{0.1cm}
    \includegraphics[width=\linewidth]{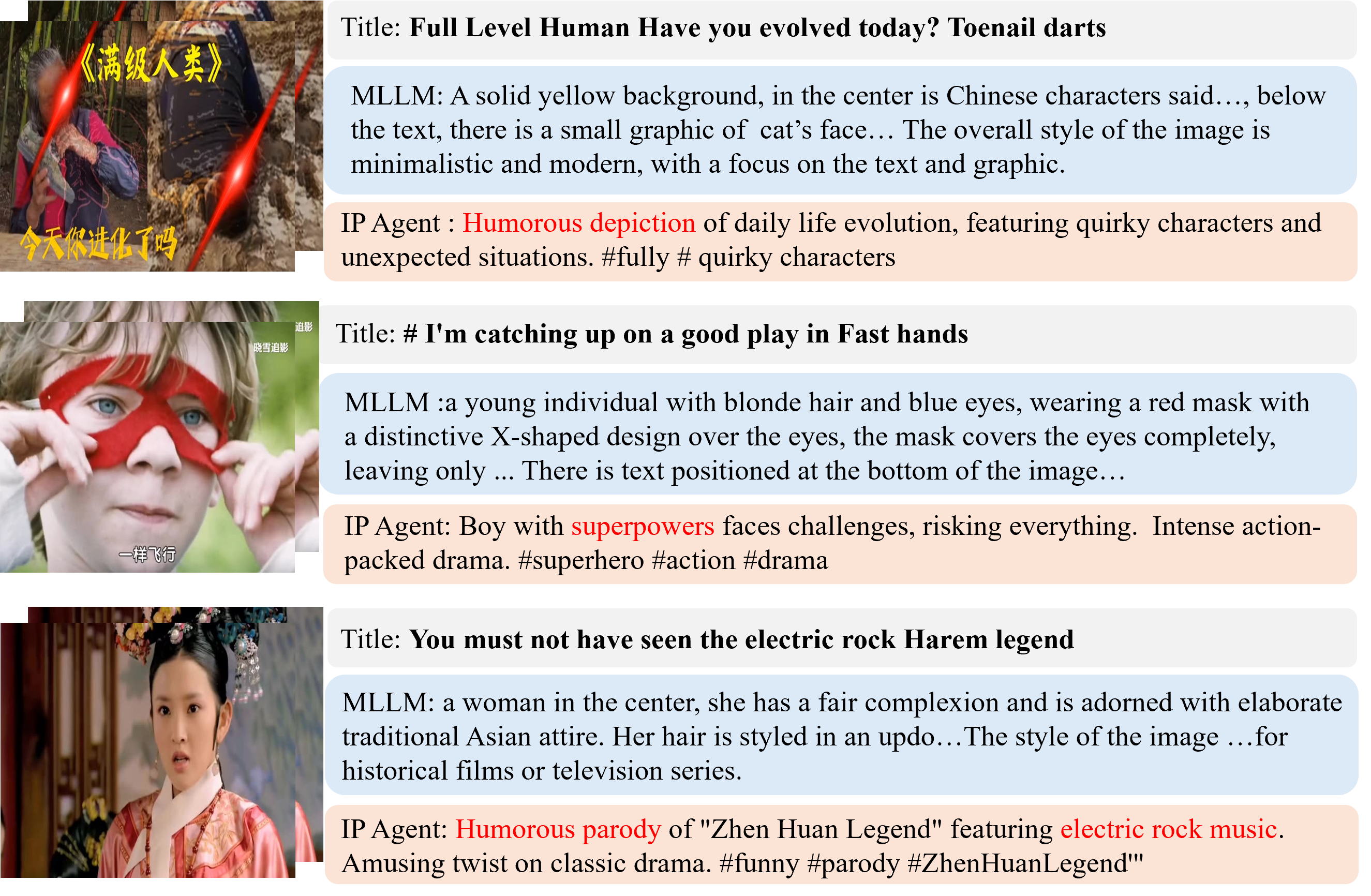}
    \caption{\textbf{Additional Visualization Results for Video Understanding.} }
    \vspace{-0.3cm}
    \label{fig:add_vis}
    \vspace{-0.2cm}
\end{figure}

In this section, we give more visualization results on the understanding of video items in Fig. \ref{fig:add_vis}.
From the above contents, it is evident that models relying solely on original video titles or unimodal visual content (MLLM) struggle to capture the high-level semantic features of videos. 
Specifically, MLLM's comprehension typically remains at the surface-level scene descriptions, failing to delve into the actual semantic relationships underlying the content. 
In contrast, 
our IP Agent framework achieves deep semantic fusion and comprehensive understanding of multimodal information by integrating the model's a priori world knowledge. 
Through systematic evaluation by a human reviewer panel, the video semantic captions generated by this framework demonstrate significantly higher semantic accuracy and content congruence compared to the outputs of unimodal models, fully illustrating the effectiveness and necessity of multimodal knowledge fusion in video semantic understanding.


\newpage
\section*{NeurIPS Paper Checklist}

\begin{enumerate}

\item {\bf Claims}
    \item[] Question: Do the main claims made in the abstract and introduction accurately reflect the paper's contributions and scope?
    \item[] Answer: \answerYes{} 
    \item[] Justification: The main claims made in the abstract and introduction accurately reflect the paper's contributions and scope of MLLM-based video recommendation. 

\item {\bf Limitations}
    \item[] Question: Does the paper discuss the limitations of the work performed by the authors?
    \item[] Answer: \answerYes{} 
    \item[] Justification: We discuss the limitations at the end of the main paper.

\item {\bf Theory assumptions and proofs}
    \item[] Question: For each theoretical result, does the paper provide the full set of assumptions and a complete (and correct) proof?
    \item[] Answer: \answerNA{} 
    \item[] Justification: We don't consider theoretical proof or result in the paper.

    \item {\bf Experimental result reproducibility}
    \item[] Question: Does the paper fully disclose all the information needed to reproduce the main experimental results of the paper to the extent that it affects the main claims and/or conclusions of the paper (regardless of whether the code and data are provided or not)?
    \item[] Answer: \answerYes{} 
    \item[] Justification:  We have fully disclose all the information in experiment details.

\item {\bf Open access to data and code}
    \item[] Question: Does the paper provide open access to the data and code, with sufficient instructions to faithfully reproduce the main experimental results, as described in supplemental material?
    \item[] Answer: \answerNo{} 
    \item[] Justification: The data is open source to access and we will provide the code after the commercial approval.  

\item {\bf Experimental setting/details}
    \item[] Question: Does the paper specify all the training and test details (e.g., data splits, hyperparameters, how they were chosen, type of optimizer, etc.) necessary to understand the results?
    \item[] Answer: \answerYes{} 
    \item[] Justification:  We specify the details in the experiments. 

\item {\bf Experiment statistical significance}
    \item[] Question: Does the paper report error bars suitably and correctly defined or other appropriate information about the statistical significance of the experiments?
    \item[] Answer: \answerYes{} 
    \item[] Justification: The main table do not provide error bars since some results are from previous papers with no error bars, and some experiments may need GPT-4o with money cost. We report the bar of our main results in the appendix.  

\item {\bf Experiments compute resources}
    \item[] Question: For each experiment, does the paper provide sufficient information on the computer resources (type of compute workers, memory, time of execution) needed to reproduce the experiments?
    \item[] Answer: \answerYes{} 
    \item[] Justification:  We use 4 80G A100 GPUs for the RFT, and the training takes about 10 hours.
    
\item {\bf Code of ethics}
    \item[] Question: Does the research conducted in the paper conform, in every respect, with the NeurIPS Code of Ethics \url{https://neurips.cc/public/EthicsGuidelines}?
    \item[] Answer: \answerYes{} 
    \item[] Justification:  We follow the NeurIPS code of Ethics.

\item {\bf Broader impacts}
    \item[] Question: Does the paper discuss both potential positive societal impacts and negative societal impacts of the work performed?
    \item[] Answer: \answerYes{} 
    \item[] Justification: Our work aims to improve video recommendation for users, enabling them to get more satisfying recommendation results. But we establish the user profile which may have privacy consideration problems.
    
\item {\bf Safeguards}
    \item[] Question: Does the paper describe safeguards that have been put in place for responsible release of data or models that have a high risk for misuse (e.g., pretrained language models, image generators, or scraped datasets)?
    \item[] Answer: \answerNA{} 
    \item[] Justification: The paper poses no such risks. 

\item {\bf Licenses for existing assets}
    \item[] Question: Are the creators or original owners of assets (e.g., code, data, models), used in the paper, properly credited and are the license and terms of use explicitly mentioned and properly respected?
    \item[] Answer: \answerYes{} 
    \item[] Justification: We cite all relevant works for the original owners of assets.

\item {\bf New assets}
    \item[] Question: Are new assets introduced in the paper well documented and is the documentation provided alongside the assets?
    \item[] Answer: \answerYes{} 
    \item[] Justification: The new assets are well documented. We prepare the documentation of our code for future reproduction and will release it afterward.  

\item {\bf Crowdsourcing and research with human subjects}
    \item[] Question: For crowdsourcing experiments and research with human subjects, does the paper include the full text of instructions given to participants and screenshots, if applicable, as well as details about compensation (if any)? 
    \item[] Answer: \answerNA{} 
    \item[] Justification: The paper does not involve crowdsourcing nor research with human subjects. 

\item {\bf Institutional review board (IRB) approvals or equivalent for research with human subjects}
    \item[] Question: Does the paper describe potential risks incurred by study participants, whether such risks were disclosed to the subjects, and whether Institutional Review Board (IRB) approvals (or an equivalent approval/review based on the requirements of your country or institution) were obtained?
    \item[] Answer: \answerNA{} 
    \item[] Justification: The paper does not involve crowdsourcing nor research with human subjects.  

\item {\bf Declaration of LLM usage}
    \item[] Question: Does the paper describe the usage of LLMs if it is an important, original, or non-standard component of the core methods in this research? Note that if the LLM is used only for writing, editing, or formatting purposes and does not impact the core methodology, scientific rigorousness, or originality of the research, declaration is not required.
    \item[] Answer: \answerYes{} 
    \item[] Justification: We use MLLM to help understand video contents and and train the LLM with reinforcement fine-tuning to simulate user decision. 

\end{enumerate}

\end{document}